\documentclass[aps,twocolumn, 10pt, superscriptaddress,showpacs,reprint, longbibliography]{revtex4-1}
\usepackage{graphicx}
\usepackage{amsmath}
\usepackage{dcolumn}
\usepackage{bm}
\usepackage{mathtools}
\usepackage{color}

\usepackage[normalem]{ulem}

\begin{document}

\preprint{APS/123-QED}

\title{Hot Particles Attract in a Cold Bath}
\author{Hidenori Tanaka}
\email{tanaka@g.harvard.edu}
\affiliation{Harvard John A. Paulson School of Engineering and Applied Sciences, Harvard University, Cambridge, MA 02138}
\affiliation{Kavli Institute for Bionano Science and Technology, Harvard University, Cambridge, MA 02138}

\author{Alpha A. Lee}
\email{alphalee@g.harvard.edu}
\affiliation{Harvard John A. Paulson School of Engineering and Applied Sciences, Harvard University, Cambridge, MA 02138}
\affiliation{Kavli Institute for Bionano Science and Technology, Harvard University, Cambridge, MA 02138}
\
\author{Michael P. Brenner}
\email{brenner@seas.harvard.edu}
\affiliation{Harvard John A. Paulson School of Engineering and Applied Sciences, Harvard University, Cambridge, MA 02138}
\affiliation{Kavli Institute for Bionano Science and Technology, Harvard University, Cambridge, MA 02138}

\date{\today}
\begin{abstract}
Controlling interactions out of thermodynamic equilibrium is crucial for designing addressable and functional self-organizing structures. These active interactions also underpin collective behavior in biological systems. Here we study a general setting of active particles in a bath of passive particles, and demonstrate a novel mechanism for long range attraction between active particles. The mechanism operates when the translational persistence length of the active particle motion is smaller than the particle diameter. In this limit, the system reduces to particles of higher diffusivity (``hot'' particles) in a bath of particles with lower diffusivity (``cold'' particles). This attractive interaction arises as a hot particle pushes cold particles away to create a large hole around itself, and the holes interact via a depletion-like attraction. 
Strikingly, the interaction range is more than an order of magnitude larger than the particle radius, well beyond the range of conventional depletion force. Although the mechanism occurs outside the parameter regime of typical biological swimmers, the mechanism could be realized in the laboratory.
\end{abstract}

\maketitle

\section{Introduction}
Active materials -- materials that comprise building blocks driven away from thermodynamic equilibrium by a constant injection of energy -- have received significant attention in recent times. Those building blocks could be biological, such as bacteria \cite{sokolov2009,leonardo2010,cates2012}, or synthetic, such as Janus colloid particles \cite{paxton2004,howse2007}. The mechanism of energy injection also varies between systems, from beating flagella to self-diffusiophoresis. Active materials serve as simple models for collective behavior in biological systems, as well as useful nanotechnological systems themselves. For example, systems can be designed to organize the incoherent motion of individual active particles to generate coherent flows \cite{stenhammar2016} and extract work \cite{thampi2016}. Underlying those disparate applications of functional active materials is the question of interparticle interactions and force generation. In equilibrium systems, complex self-assembled structures can be engineered by tuning interactions between components \cite{nykypanchuk2008dna,ke2012three}. However, concepts such as the potential of mean force in equilibrium systems cannot be readily extended to nonequilibrium systems. Moreover, active materials, unlike equilibrium materials, have the ability to continuously do work and thus could open up new regions in material design space \cite{ding2016light,martinez2016colloidal}. Nonetheless, the mechanisms of active force generation are yet to be fully understood even at the phenomenological level. 

A simple and ubiquitous setting in active materials is active particles with energy input into translation and rotation degrees of freedom embedded in a bath of passive particles \cite{stenhammar2015activity}. Some examples include biofilms which are composed of both living and dead bacteria \cite{chai2011,drescher2011}, cell migration through passive tissues \cite{trepat2009,angelini2011}, sperms swimming through the viscoelastic cervical mucus \cite{suarez2005,tung2015microgrooves}, chromosome positioning in eukaryotic cell nuclei \cite{ganai2014chromosome}, passive tracer particles being stirred by the motion of active bacteria or algae \cite{wu2000particle,chen2007fluctuations,leptos2009dynamics,valeriani2011colloids,mino2011enhanced,mino2013induced}, crystallization of colloidal systems promoted by doping with self-propelling particles \cite{ni2014crystallizing,kummel2015formation}, and rotating active particles that interact via hydrodynamic forces between vortices \cite{grzybowski2000dynamic,goto2015purely} coupled with surrounding passive particles \cite{aragones2016elasticity,steimel2016emergent}. 
Furthermore, interactions between passive particles in a bath of active particles, commonly referred to as ``active depletion force'', have been studied in various settings \cite{ni2015tunable, harder2014role, smallenburg2015swim}.

Naturally, the majority of studies to date focus on active particles with translational persistence length longer than their diameter, such as self-propelling particles with translation and rotation degrees of freedom coupled via the Stokes-Einstein relation
\cite{angelani2011effective,valeriani2011colloids,mccandlish2012spontaneous,fily2012cooperative,ni2013pushing,ni2014crystallizing,das2014phase,stenhammar2015activity,bialke2015negative} 
in an effort to understand collective phenomena of biological swimmers.
However, with the advent of many artificial swimmer systems and self-propelling structures \cite{maass2016,bechinger2016active}, reaching the parameter space outside biomimicry opens up a rich design space of self-organizing structures.  

In this paper, we show that a \emph{long-ranged} attraction (up to an order of magnitude larger than the particle diameter) between active particles in a passive bath can be engineered when the translational persistence length of the active particle is shorter than the particle diameter. In this limit, the system reduces to Brownian particles of higher diffusivity (``hot'' particles) in a bath of Brownian particles of lower diffusivity (``cold'' particles). 
The hot particles attract in a cold bath because the hot particle pushes cold particles away to create a large hole around itself, and the density gradient mediates enhanced diffusion of a hot particle in the direction of the other hot particle. This mechanism is qualitatively different from recent studies \cite{weber2016binary,grosberg2015nonequilibrium} of short-ranged attraction between cold particles in a hot bath.

\begin{figure}
\begin{center}
\includegraphics[width=\columnwidth]{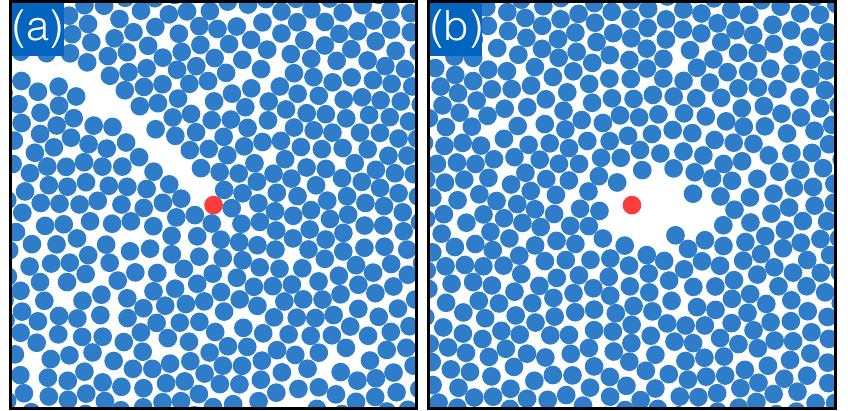}
 \caption{ Snapshots from Brownian dynamics simulations of a run-and-tumble particle (red) immersed in dense passive particles (blue) with area fraction $\phi = 0.68$. 
(a) When translational persistence length is longer than the particle diameter ($v_0 \tau=100a>a$), the motion of the run-and-tumble particle creates a void tail region where the passive particles are wiped out. 
(b) A circular excluded area is created around the particle when the tumbling rate $1/\tau$ is increased so that the translational persistence length is of the order of its diameter ($v_0 \tau=a$). }
\label{figure1}
\end{center}
\end{figure}

\section{The effect of translational persistence length: tail vs hole}
%\subsection{Run-and-tumble model}
To fix ideas, we first consider a single run-and-tumble particle, which resembles the motion of bacteria such as \emph{Escherichia coli} \cite{berg1972chemotaxis}, immersed in a bath of dense passive particles. The system consists of $N=1600$ particles having area fraction $\phi=0.68$ and the dynamics of $i$th particle at position vector $\bm{r_{i}}$ satisfies 
\begin{equation}
\partial_{t} \bm{ r}_{i} = v_{i}\bm{\hat{v}_{i}}(\theta_i) - \mu \sum_{j} { \nabla } U(r_{ij}) + \bm{ \eta }_{i} (t)
\label{SPPeq}
\end{equation}
where $v_{i}$ is the swimming velocity, $\bm{\hat{v}_{i}}(\theta_i)=(\cos\theta_i (t), \sin\theta_i (t) )$ is orientation of $i$th particle and $\mu$ is the mobility. We set $v_{i}=0$ for passive particles and $v_{i}=v_{0}$ for the run and tumble particle. 
The orientation of a run-and-tumble particle, $\theta_i$, is randomly chosen from $[-\pi, \pi]$, every $\tau a^2/D$ time interval, where $a$ is particle's radius, and $D$ is the diffusion constant.
The stochastic terms describe thermal noise from solvent molecules and satisfy
$\big \langle \eta_{i\alpha}(t)\eta_{j\beta}(t') \big \rangle = 2D\delta_{ij} \delta_{\alpha \beta} \delta(t-t')$, where $\alpha$ and $\beta$ denote the spatial components.
To capture excluded volume effects, we include a repulsive truncated harmonic potential between the particles $U(r_{ij})=\frac{1}{2}k (2a-|\bm{r}_i - \bm{r}_j|)_{+}^2$ with spring constant $k=100 D/\mu a^2$.
Whilst the parameters of the biological and synthetic run-and-tumble particles vary among different systems \cite{bechinger2016active}, we fix $v_0=1000D/a$ which corresponds to: $v_0=100\mu m /s$, $D=0.1\mu m^2/s$ and $a=1\mu m$. This swim speed is the typical order of magnitude for systems such as spermatozoa and \emph{T. majus}, as well as many synthetic swimmer systems.
The time interval between tumbling, $\tau a^2/D$, is the key control parameter of the system.
Fig.~\ref{figure1}(a) shows that the run-and-tumble particle wipes out passive particles to leave a tail of excluded region behind its trajectory when the persistence length of the swimming motion is larger than the particle radius ($v_0 \tau=100a>a$); we note that typical biological swimmers are elongated, and for simplicity we focus on the spherical particles which is an appropriate model for systems such as self-propelled Janus colloidal particles. 

However, when the persistence length is much shorter than the typical bacterium and comparable to the particle radius, the diffusive motion of the run-and-tumble particle isotropically pushes out the passive particles and creates a circular excluded area (Fig.~\ref{figure1}(b)). Considering length scales larger than the persistence length of the particle's translational motion, the effective diffusion constant of a run-and-tumble particle is $D_{\mathrm{eff}} \sim v_0^2 \tau$ \cite{PhysRevLett.100.218103}, which is much larger than the diffusion coefficient of the surrounding passive particles. In other words, in the limit where the persistence length of the run-and-tumble particle is comparable to the particle radius, the phenomenology of the system can be reduced to that of a ``hot particle'' in a ``cold'' bath \cite{weber2016binary}.  Appendix A shows that similar phenomenology is seen for the self-propelling particle model. 

In the remainder of this paper, we will focus on the phenomenology of ``hot'' particles with diffusion coefficient $D_{\rm{hot}}$ in a bath of ``cold'' particles with $D_{\rm{cold}}$. The control parameters of the system are ratio of the diffusion constants $D_{\rm{hot}}/D_{\rm{cold}}$ and the area fraction $\phi$.

\begin{figure}[!tbp]
\begin{center}
\includegraphics[width=\columnwidth]{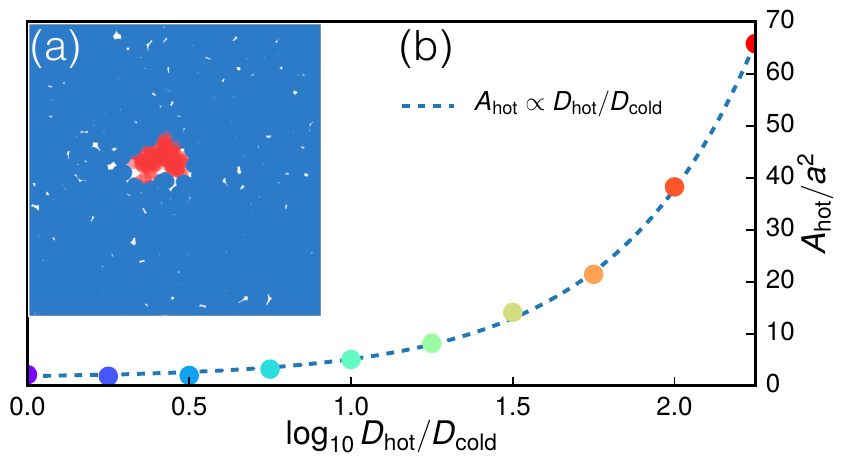}
\caption{
(a) A hot particle (red) creates a circular hole around itself when immersed in dense cold particles (blue). With $\log_{10} D_{\rm{hot}}/D_{\rm{cold}}=2.0$, traces of the hot and cold particles are plotted every $0.4 a^2/D_{\rm{hot}}$ time step for total time duration $40 a^2/D_{\rm{hot}}$. (b) The size of the hole increases linearly as the temperature difference between the hot particle and its environment increases. For both plots, the area fraction of the system are set to be $\phi=0.68$. }
\label{figure2}
\end{center}
\end{figure}
 
Fig.~\ref{figure2}(a) shows that the two temperature system qualitatively reproduces the circular excluded area observed in Fig.~\ref{figure1}(b). 
Tracing the trajectory of hot and cold particles, Fig.~\ref{figure2}(a) shows that the hot particle has a much larger effective area than the cold particles because of its high diffusivity. The hole size $A_{\rm{hot}}$ is linearly proportional to $D_{\rm{hot}}/D_{\rm{cold}}$ (Fig.~\ref{figure2}(b)). (Appendix B describes how $A_{\rm{hot}}$ is estimated.)
This scaling can be shown using a simple pressure-volume argument: 
First, we note that the mobility $\mu$ of both hot and cold particles are the same and the relation $\mu = D_{\rm{hot}}/k_{\rm{B}}T_{\rm{hot}} = D_{\rm{cold}}/k_{\rm{B}}T_{\rm{cold}}$ holds.
In the limit where the system size is much larger than the area created by single hot particle, the compressibility of cold bath is constant. 
Therefore, the pressure of cold bath $p_{\rm{cold}}$ can be related to the area occupied by the cold bath $A_{\rm{cold}}$ and diffusion constant (i.e. temperature) $D_{\rm{cold}}$ via $p_{\rm{cold}} \propto D_{\rm{cold}}/A_{\rm{cold}}$. 
Similarly, the hot particle exerts a pressure $p_{\rm{hot}} \propto D_{\rm{hot}}/A_{\rm{hot}}$. By equating the pressures, we arrive at $A_{\rm{hot}}/A_{\rm{cold}} = A_{\rm{hot}}/(A - A_{\rm{hot}}) \propto D_{\rm{hot}}/D_{\rm{cold}}$, where $A$ is the system area. As $A_{\rm{hot}} \ll A$, we expect a linear relationship 
\begin{equation}
A_{\rm{hot}} \propto D_{\rm{hot}}/D_{\rm{cold}}
\end{equation}
which is indeed seen in Fig.~\ref{figure2} (b).

\begin{figure}[!tbp]
\begin{center}
\includegraphics[width=\columnwidth]{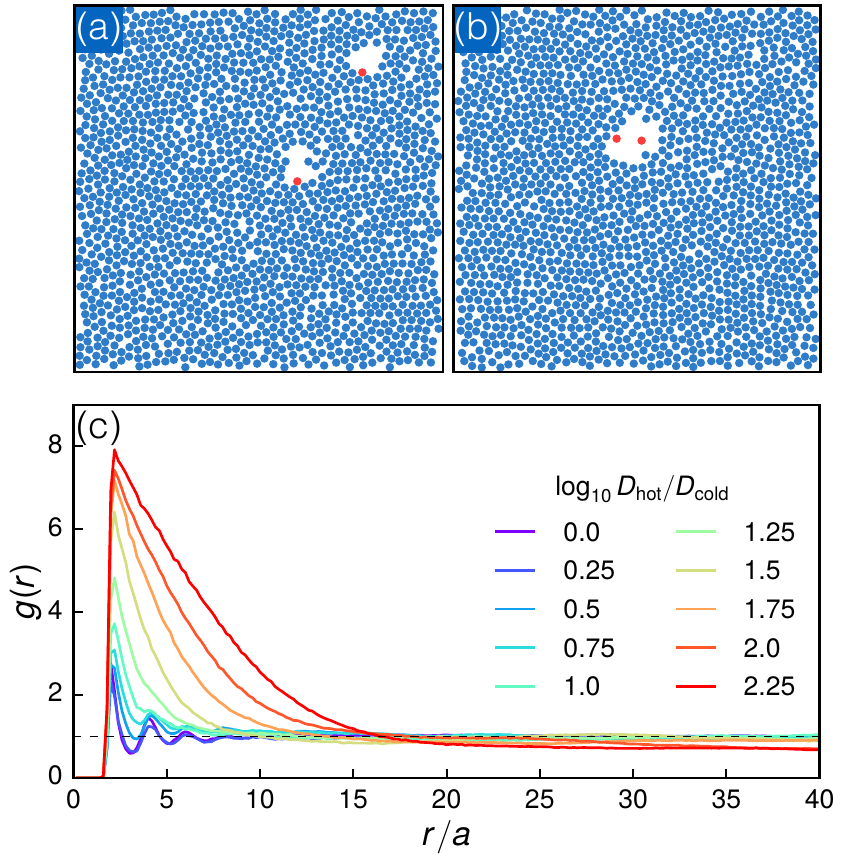}
\caption{ Snapshots from a simulation with $\log_{10} D_{\rm{hot}}/D_{\rm{cold}}=2$: (a) Each of the two hot particles creates hole around them. (b) Two hot particles are caged once they are close enough. (c) Radial pair distribution function $g(r)$ between two hot particles immersed in dense cold particles with various diffusivity ratio $D_{\rm{hot}}/D_{\rm{cold}}$. Area fraction of the system is set to be $\phi=0.68$.} 
\label{figure3}
\end{center}
\end{figure}

\section{Emergence of long-range force between hot particles in a cold bath}

Thus far, we studied the density field of cold particles around a single hot particle and showed the formation of a circular excluded area in the non-equilibrium steady state. To study the inter-particle force between two hot particles, we simulate a system of $N=1600$ particles with area fraction $\phi=0.68$ starting from random initial conditions.
Fig.~\ref{figure3} shows that such density inhomogeneity generates a force between the hot particles. One may naively expect the two hot particles to quickly run away from each other due to enhanced diffusion, which is the case without surrounding passive particles. However, our simulation shows that two initially separated hot particles independently create holes around themselves and the holes diffuse in the system (Fig.~\ref{figure3} (a)). Once the two holes overlap, the two hot particles are caged in a larger hole, thus creating an effective attraction. 

This interaction can be quantified by the radial pair distribution function $g(r)$ between two hot particles immersed in dense cold particles. $g(r)$ is numerically estimated by running 11 independent simulations over $1.6\times 10^{6} D_{\rm{hot}}/a^2$ time duration in total while recording distance between the two hot particles every $0.2 D_{\rm{hot}}/a^2$, for each diffusivity ratio $D_{\rm{hot}}/D_{\rm{cold}}$. When $D_{\rm{hot}}/D_{\rm{cold}}=1$, the radial distribution function is that of a simple liquid, showing small oscillations reflecting local structural ordering due to volume exclusion. As $D_{\rm{hot}}/D_{\rm{cold}}$ increases, the radial distribution function shows a clear peak indicating the existence of an attractive interaction between the two hot particles (Fig.~\ref{figure3} (c)).  A large $D_{\rm{hot}}/D_{\rm{cold}}$ is required to generate a strong and long-ranged force between the hot particles.

\begin{figure}[!tbp]
\begin{center}
\includegraphics[width=\columnwidth]{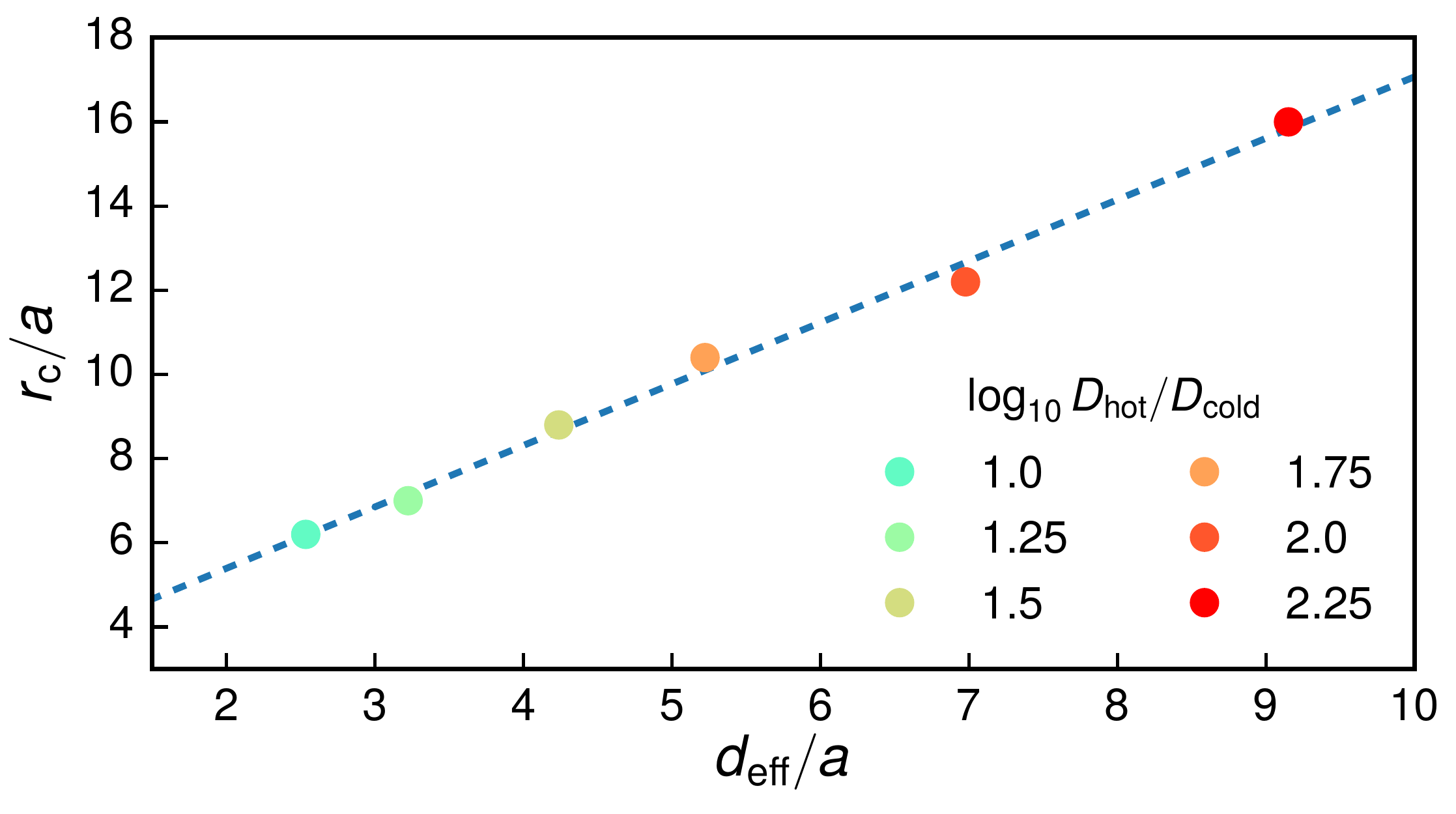}
\caption{ The interaction range $r_{\rm{c}}$ of the caging force, calculated from $g(r)$ (shown in Fig.~\ref{figure3}), is directly proportional to the diameter of a hole $d_{\rm{eff}}$ (shown in Fig.~\ref{figure2}). Dashed line is the result of a linear fit.}
\label{figure4}
\end{center}
\end{figure}

The attractive interaction is tunable by controlling the hole size around a single hot particle, which in turn can be varied by changing the diffusivity ratio between the hot particles and the surrounding cold particles (c.f. Fig.~\ref{figure2}). Fig.~\ref{figure4} shows that the diameter of the excluded hole is proportional to the range of the force. The effective diameter of a hole $d_{\rm{eff}}$ is calculated from the measured area $A_{\rm{hot}}$ assuming circular geometry $d_{\rm{eff}}=2\sqrt{A_{\rm{hot}}/\pi}$. To obtain the interaction range, we define an effective interaction potential \cite{PhysRevLett.113.238701}.
\begin{equation}
U_{\rm{eff}}(r) \propto - \log g(r) , 
\end{equation}
and the $F_{\rm{eff}}(r)=-\mathrm{d}U_{\rm{eff}}/\mathrm{d}r$. We take the effective interaction range $r_{\rm{c}}$ to be the largest $r$ that satisfy $F_{\rm{eff}} > -0.1$.

\begin{figure}[!tbp]
\begin{center}
\includegraphics[width=\columnwidth]{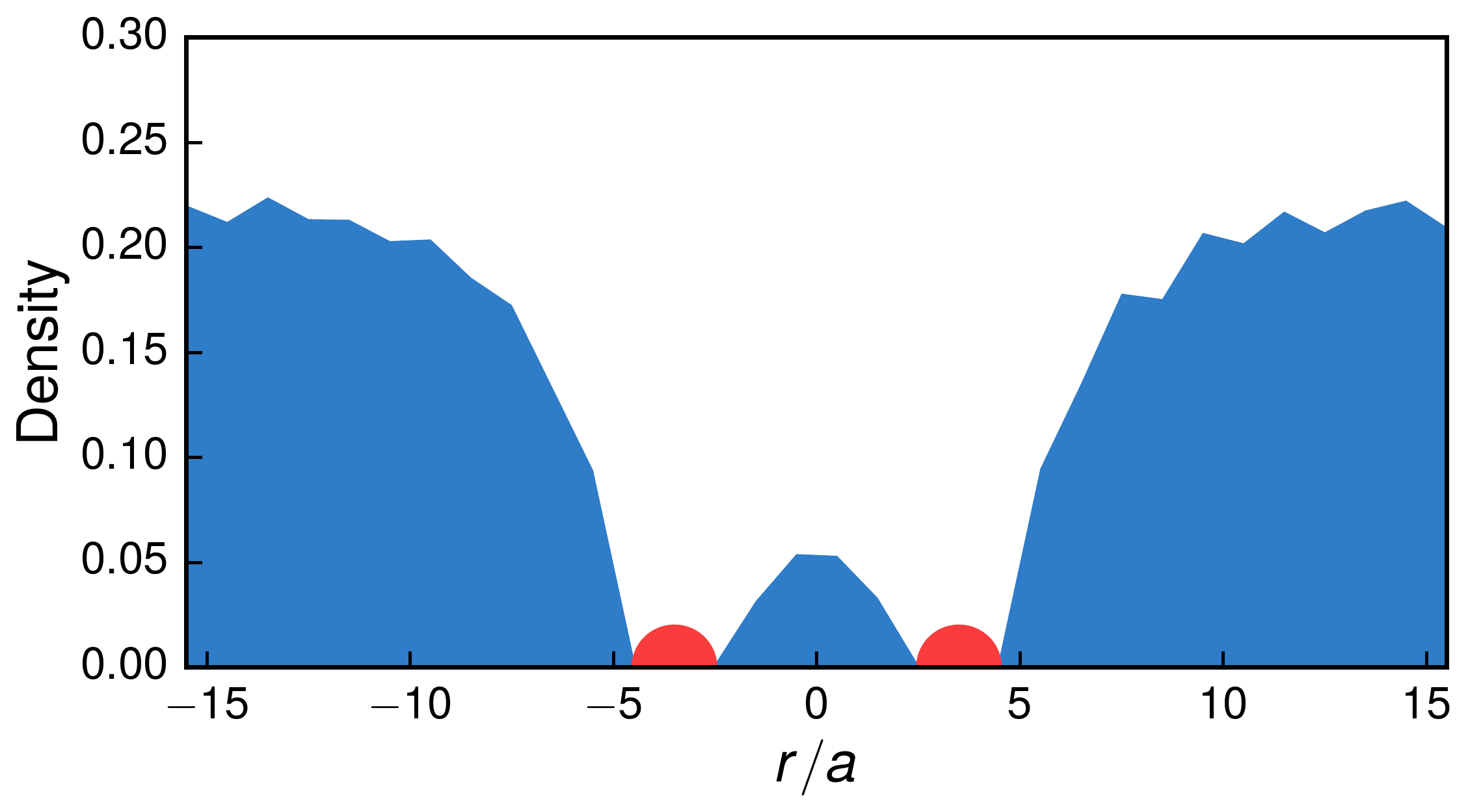}
\caption{Time averaged number density profile of cold particles along two hot particles separated by $6 a < r <8 a$ with $\phi=0.68$ and $\log_{10} D_{\rm{hot}}/D_{\rm{cold}} =2.0$. The density of cold particles is significantly reduced in-between the hot particles, resulting in a net diffusion in direction of the other hot particle and thus an attractive interaction. 
}
\label{figure5}
\end{center}
\end{figure}

\begin{figure}[!tbp]
\begin{center}
\includegraphics[width=\columnwidth]{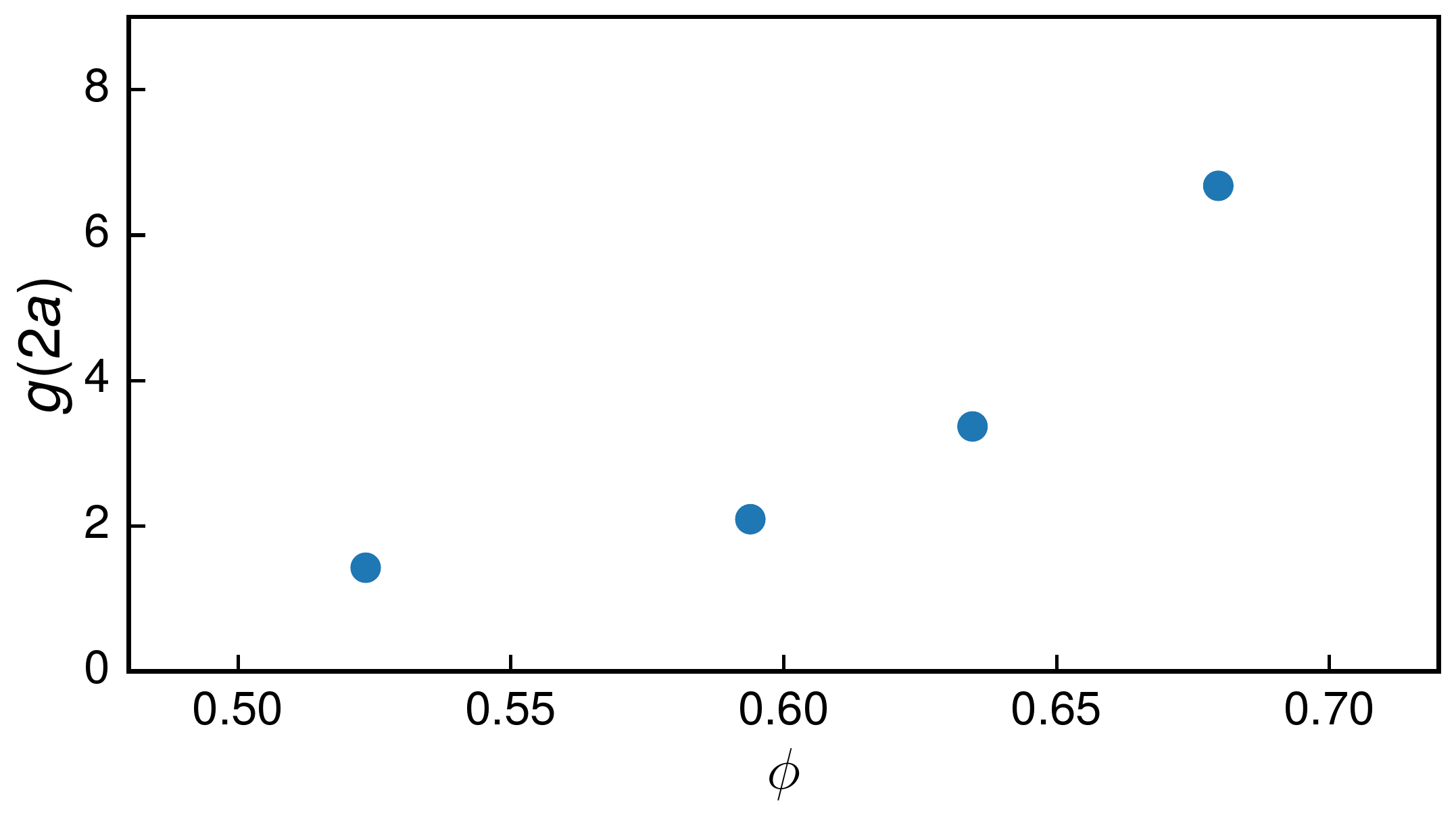}
\caption{The pair distribution distribution function of the hot particles at contact, $g(2 a)$, as a function of packing fraction of cold particles $\phi$. The diffusivity ratio is taken to be $D_{\mathrm{hot}}/D_{\mathrm{cold}} = 100$, and the number of cold particles is $N=1600$. }
\label{figure6}
\end{center}
\end{figure}

The microscopic mechanism of the attractive interaction and caging effect is shown in Fig.~\ref{figure5}. 
For each snapshot of simulation in which the two hot particles are separated by $6a<r<8a$, we calculate density profile of cold particles along a line connecting the two hot particles and averaged the profiles over time.
There are less cold particles in-between the hot particles than in the bulk. The density gradient in cold particles generates a force as the hot particles experience less collision by diffusing towards each other than away from each other, analogous to an osmotic force at equilibrium. In particular, the fact that the range of the force $r_{\rm{c}} \propto d_{\rm{eff}}$ is reminiscent of the well-known depletion interaction in equilibrium \cite{asakura1954interaction}. The equilibrium depletion interaction is caused by the exclusion of smaller particles in the gap between larger particles; here, although the size of the hot particles is the same as that of the cold particles, the excluded hole around the hot particle interacts as if it were a larger particle. The excluded hole around the hot particle becomes ``softer'' and more deformable as the packing fraction of the cold bath decreases, thus a corollary of the depletion interaction analogy is that the strength of the interaction should decrease as  the packing fraction of the cold bath decreases; this is indeed confirmed in Fig. \ref{figure6}. We could rule out thermophoresis as the gradient in mean square velocity of the particles in the colder bath surrounding the hot particle is negligible, as one would expect from an overdamped system (See Appendix C). 

\section{Conclusion}
In summary, we found an attractive interaction between highly diffusive particles immersed in a bath of dense and less diffusive particles of the same size. The key ingredient of this force is a translational persistence length smaller than the particle radius, which requires engineering the tumbling rate or rotational diffusion coefficient such that it is much larger than the typical bacterium or Janus swimmer. Formation of a circular excluded area around the highly diffusive particles is responsible for the attractive interaction, and both the range as well as the strength of the interaction can be tuned by controlling activity and density. Crucially, the interaction range can be much longer than the particle's radius ($>10a$) depending on characteristic size of a hole caging the two particles. 

We note that the mechanism behind the attractive interaction between two hot particles is fundamentally different from the so-called active depletion force \cite{harder2014role, smallenburg2015swim}. First, the active depletion force between large passive objects immersed in a bath of smaller swimmers is repulsive. Second, mechanistically the active depletion force arises due to directional correlation of the swimming velocity field which creates a large swim pressure between the passive particles. The phenomenology of hot particles attracting in a cold bath is also distinct from the clustering of cold particles in a bath of hot particles reported recently \cite{weber2016binary}. The interaction range of the force reported in \cite{weber2016binary} is short-ranged and decays over the length scale of a particle, whereas the force reported here has a range that is an order of magnitude larger than the particle diameter. 

Mixture of self-propelled particles in dense passive media is experimentally realizable \cite{bechinger2016active}. By identifying the translational persistence length as the key control parameter for force generation, we hope our results will inspire experimental studies in tuning the persistence length. For artificial swimmers, the persistence length could be decreased by strategies such as putting geometric anisotropy in particle shape \cite{kummel2013circular} or applying an external alternating field \cite{mano2016optimal}. We also note that a macroscale prototype of our effect may be realized by shaking granular mixture of grains with different friction coefficients/masses \cite{grunwald2016exploiting}.

\acknowledgments
H.T. would like to thank K. Kawaguchi for fruitful discussions and comments on manuscript. H.T. is supported by the Funai Foundation for Information Technology through Funai Overseas Scholarship. A.A.L. is supported by a Fulbright Fellowship and a George F. Carrier Fellowship at Harvard University. M.P.B. is an investigator of the Simons Foundation.
This research was funded by the National Science Foundation through Grant DMR-1435964, the Harvard Materials Research Science and Engineering Center Grant DMR1420570, and the Division of Mathematical Sciences Grant DMS- 1411694. The computations in this paper were run on the Odyssey cluster supported by the FAS Division of Science, Research Computing Group at Harvard University. 

\appendix

\section{Self-propelling particles in a bath of passive particles}
In the main text, we discussed the relationship between run-and-tumble particles and two temperature systems. In this section, we will show the limit in which self-propelling particles immersed in a bath of passive particles also reduces to a two temperature system. We consider a system of $N=1600$ particles having area fraction $\phi=0.68$ and the dynamics of $i$th particle at position vector $\bm{r_{i}}$ satisfies 
\begin{equation}
\partial_{t} \bm{ r}_{i} = v_{i}\bm{\hat{v}_{i}} - \mu \sum_{j} { \nabla } U(r_{ij}) + \bm{ \eta }^{\rm{T}}_{i} (t)
\label{SPPeq1}
\end{equation}
where $v_{i}$ is the self-propulsion velocity, $\bm{\hat{v}_{i}}(\theta_i)=(\cos\theta_i (t), \sin\theta_i (t))$ is orientation of $i$th particle and $\mu$ is the mobility. We set $v_{i}=0$ for passive particles and $v_{i}=v_{0}$ for the active particle. The stochastic terms describe random collisions with solution molecules and satisfy
$\big \langle \eta^{\rm{T}}_{i\alpha}(t)\eta^{\rm{T}}_{j\beta}(t')  \big \rangle =   2D_{\rm{T}}\delta_{ij} \delta_{\alpha \beta} \delta(t-t')$, where $\alpha$ and $\beta$ represents spatial components.
The rotational diffusion of particle's orientation is given by  
\begin{equation}
\partial_{t} \theta_i = \eta_{i}^{\rm{R}} (t)
\end{equation}
and stochastic term $\eta_{i}^{\rm{R}} (t)$ satisfies
$\big \langle \eta^{\rm{R}}_{i}(t)\eta^{\rm{R}}_{j}(t')  \big \rangle =  2D_{\rm{R}}\delta_{ij} \delta(t-t')$. To mimic hard core repulsion, we use a harmonic potential $U(r_{ij})=\frac{1}{2}k (r_{ij}-2a)_{+}^2$ with spring constant $k=100 D_{\rm{T}}/\mu a^2$ and cutoff at $r_{ij}=2a$, where $a$ is radius of a particle.

\begin{figure}
\begin{center}
\includegraphics[width=\columnwidth]{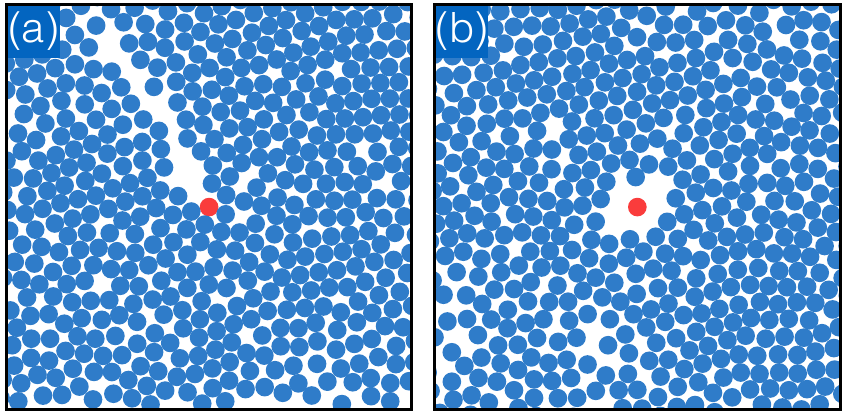}
    \caption{ Snapshots from Brownian dynamics simulations of a self-propelled particle (red) immersed in dense passive particles (blue) with area fraction $\phi = 0.68$. (a) When rotation and translation diffusion of the self-propelled particle are coupled to satisfy the Stokes-Einstein relationship $D_{\rm{R}}=3D_{\rm{T}}/a^2$, the motion of the active particle creates void tail region where the passive particles are wiped out. (b) When the rotation and translation diffusion of the active particle are uncoupled so that the persistence length of self-propelled motion is shorter than its diameter, a circular excluded area is created around the particle. }
\label{figureSI1}
\end{center}
\end{figure}

Fig.~\ref{figureSI1}(a) shows that the self-propelled particle wipes passive particles out of its trajectory when self-propelling velocity $v_{0}=1000 D_{\rm{T}}/a$ is sufficiently large compared to the translational diffusion constant. For self-propelling particle whose translational and rotational diffusion constants satisfy the Stokes-Einstein relationship $D_{\rm{R}}=3D_{\rm{T}}/(2a)^2$, the self-propelled particle shows directional motion with a persistence length much larger than its radius. Due to the long persistence length, a self-propelled particle creates an elongated void tail region along its trajectory (Fig.~\ref{figureSI1}(a)). The shape of the void region varies with different $v_{0}$ and $D_{\rm{T}}$, but it takes a circular bubble-like shape when the persistence length of the self-propelled particle is smaller than its radius. Fig.~\ref{figureSI1}(b) shows an almost circular void region when $v_{0}=1000 D_{\rm{T}}/a$ and $D_{\rm{R}}=1000 D_{\rm{T}}/a^2$.

In the limit $D_{\rm{R}} a/v_0 \rightarrow \infty$, the system reduces to a two component system with the active (passive) particle having higher (lower) diffusivity -- a two temperature system \cite{grosberg2015nonequilibrium,weber2016binary}. This asymptotic reduction can be seen by integrating over the angular dynamics and transforming Equation (\ref{SPPeq}) into a Langevin equation with a non-Markovian noise \cite{fily2012athermal}
\begin{equation}
\partial_{t} \bm{ r}_{i} = \mu \sum_{j} - { \nabla } U(r_{ij}) + \bm{ \eta }_{i} (t),
\end{equation}
with 
\begin{equation}
 \left< \eta_{i\alpha}(t)\eta_{j\beta}(t')  \right> =   \left(2D_{\rm{T}} \delta(t-t') + \frac{v_0^2}{2} e^{- D_{\rm{R}} |t-t'|} \right) \delta_{ij} \delta_{\alpha \beta} 
\label{noise}
\end{equation}
In the limit where the rotational persistence time $D_{\rm{R}}^{-1}$ is much less than the translational persistence time $a/v_0$, the exponential in (\ref{noise}) can be approximated by a $\delta$ function, giving an effective diffusion coefficient $D_{\rm{eff}} = 2 D_T + v_0^2/(2 D_{\rm{R}})$.

\section{Estimating the hole size}
\begin{figure}[!tbp]
\begin {center}
\includegraphics[width = \columnwidth]{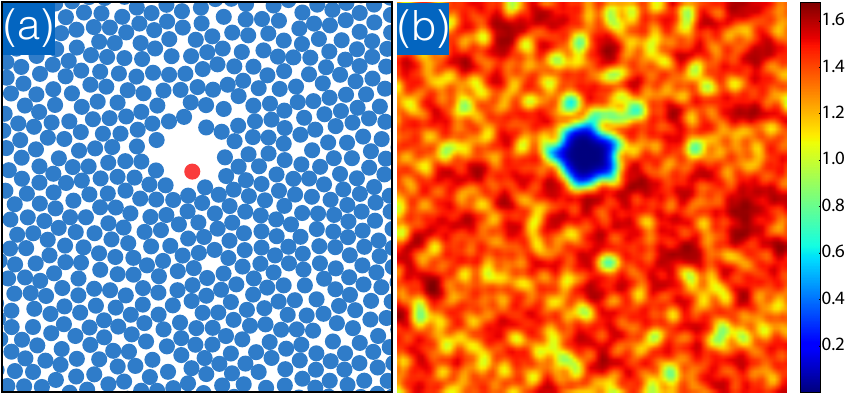} 
\caption{(a) A snapshot of a hole created around single hot particle. (b) By superposing Gaussian functions with height $1$ at each coordinate of a cold particle, we can obtain continuous density field represented as the color bar. In the snapshot, we set $D_{\rm{hot}}/D_{\rm{cold}}=10^2$ and $\phi = 0.68$.
}

\label{figureSI2}
\end{center}
\end{figure}

In this section we describe how we estimate the area of a hole from each snapshot of the system.
Given the spatial coordinate of each cold particle $(x_i,y_i)$, as shown in Fig.~\ref{figureSI2}(a), we put a Gaussian function with height $1$ at each position and superpose the Gaussian functions over all cold particles   
\begin{equation}
G(x,y)= \sum_{i=1, \neq \rm{hot}}^{N} e^{- ((x-x_i)^2 +(y-y_i)^2)/2}
\end{equation}
to define continuous density field $G(x,y)$ as shown in Fig.~\ref{figureSI2}(b).
We can then map the continuous function $G(x,y)$ to a binary-valued function $\bar{G}(x,y)$ by defining a threshold, which is chosen to be $0.5$ for our analysis.

\section{The absence of a gradient in $\langle v^2 \rangle$ around a single hot particle}
\begin{figure}[!ht]
\begin {center}
\includegraphics[width = \columnwidth]{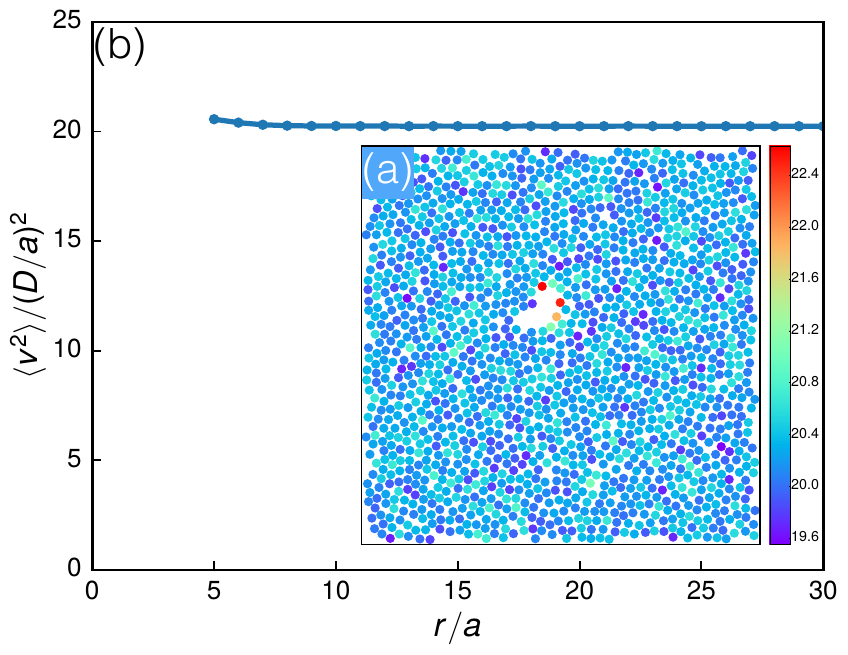} 
\caption{ (a) A snapshot of cold particles. Each particle is colored based on $\langle v^2 \rangle D_{\rm{hot}}/a$ as specified in the color bar. The simulations are done with parameters $D_{\rm{hot}}/D_{\rm{cold}}=10^{2}$ and $\phi=0.68$, and the time average is taken for $20 a^2/D_{\rm{hot}}$.
(b) $\langle v^2(r) \rangle$ is plotted as a function of radial distance from center of a hole created around a hot particle.
}

\label{figureSI4}
\end{center}
\end{figure}

In this paper, we considered Brownian dynamics and neglected the inertial term to focus on the role of diffusivity. We argue that a gradient in the density of cold particle is the microscopic mechanism of force generation. One might wonder whether a gradient in mean squared velocity, in other words ``temperature'', is a confounding explanation for this force. Fig.~\ref{figureSI4} shows that force generation is not due to a gradient in $\langle v^2 \rangle$, thus thermophoresis is not the mechanism of force generation.

From our numerical results, $v^2$ is calculated as 
\begin{equation}
v^2 \equiv \frac{ ( x(t+\Delta t) - x(t) )^2 + ( y(t+\Delta t) - y(t) )^2 }{\Delta t^2}
\end{equation}
with $\Delta t=0.002 a^2/D$.

A snapshot of the system is shown in Fig.~\ref{figureSI4}(a). We color each particle according to $v^2$ averaged over time $20 a^2/D_{\rm{hot}}$, which is smaller than characteristic translational time scale for cold particles $a^2/D_{\rm{cold}}$, and set parameters $D_{\rm{hot}}/D_{\rm{cold}}=10^{2}$ and $\phi=0.68$. The snapshot shows that while a few particles at the contour of a hole have a slightly higher $v^2$, a pronounced gradient is absent.

To further quantify the profile of the mean-squared velocity around a single hot particle, Fig~\ref{figureSI4}(b) shows $\langle v (r)^2  \rangle$, which is $v^2$ time averaged over $20000a^2/D_{\rm{hot}}$, where $r$ is the distance from center of area of the hole. It is clear that $\langle v (r)^2  \rangle$ is almost constant, as expected from the overdamped dynamics of the system.

\cleardoublepage
\bibliography{reference,library}
\end{document}